\documentclass[10pt, conference]{IEEEtran}
\usepackage[utf8]{inputenc}
\usepackage{cite}
\usepackage{amsmath,amssymb,amsfonts}
\usepackage{algorithmic}
\usepackage{graphicx}
\usepackage{textcomp}

\usepackage[dvipsnames]{xcolor}
\usepackage{url}
\usepackage{hyperref}
\usepackage{booktabs}

\usepackage{soul} 
\usepackage{comment}

\newcommand{\method}{Differentiable Dictionary Search}
\newcommand{\abbrev}{DDS}

\def\BibTeX{{\rm B\kern-.05em{\sc i\kern-.025em b}\kern-.08em
    T\kern-.1667em\lower.7ex\hbox{E}\kern-.125emX}}
\begin{document}

\title{Probabilistic Modelling of Signal Mixtures \\
with Differentiable Dictionaries
\\
\thanks{The LIT AI Lab
is supported by the Federal State of Upper Austria.}}

\author{\IEEEauthorblockN{Lukas Martak}
\IEEEauthorblockA{\textit{CP JKU / LIT} \\
\textit{Johannes Kepler University}\\
Linz, Austria \\
lukas.martak@jku.at}
\and
\IEEEauthorblockN{Rainer Kelz}
\IEEEauthorblockA{\textit{CP JKU} \\
\textit{Johannes Kepler University}\\
Linz, Austria \\
rainer.kelz@jku.at}
\and
\IEEEauthorblockN{Gerhard Widmer}
\IEEEauthorblockA{\textit{CP JKU / LIT} \\
\textit{Johannes Kepler University}\\
Linz, Austria \\
gerhard.widmer@jku.at}
}

\author{\IEEEauthorblockN{Luk\'a\v{s} Samuel Mart\'ak$^{1,2}$, Rainer Kelz$^{1}$, and Gerhard Widmer$^{1,2}$}
\IEEEauthorblockA{$^1$Institute of Computational Perception \& $^2$LIT Artificial Intelligence Lab \\
Johannes Kepler University Linz, Austria \\
\{first.last\}@jku.at}
}

\maketitle
\thispagestyle{plain}
\pagestyle{plain}

\begin{abstract}
We introduce a novel way to incorporate prior information into (semi-) supervised non-negative matrix factorization, which we call \textit{differentiable dictionary search}. It enables general, highly flexible and principled modelling of mixtures where non-linear sources are linearly mixed. We study its behavior on an audio decomposition task, and conduct an extensive, highly controlled study of its modelling capabilities.
\end{abstract}


\section{Introduction}

Non-negative matrix factorization (NMF) \cite{Lee1999LearningTP} can be used to decompose a spectrogram $\mathbf{S}$ of an audio mixture into a spectral representation of its individual sources \cite{DBLP:conf/ica/Smaragdis04}. In its unsupervised form, NMF simultaneously tries to learn a representation of the individual sources, a \textit{dictionary} $\mathbf{W}$, as well as the points in time at which those sources can be heard --- their \textit{activity} over time $\mathbf{H}$, such that $\mathbf{S} \approx \mathbf{W}\mathbf{H}$. This decomposition succeeds if one crucial assumption about the audio mixture is met: all the individual sound sources that are mixed together can also be heard in isolation for at least some amount of time. If this assumption is not true, NMF will not be able to separate sound sources that only appear together \cite{DBLP:conf/nips/DonohoS03}. To still achieve a useful decomposition in such cases, some prior information about the individual sound sources needs to be incorporated, for example as structural constraints on the dictionary of sources, the vectors in $\mathbf{W}$ that describe the spectral representation of the sound sources. This approach is called \textit{supervised} or \textit{semi-supervised} NMF \cite{DBLP:conf/ica/SmaragdisRS07} or alternatively \textit{task-driven dictionary learning} \cite{DBLP:journals/pami/MairalBP12}. For this to work we need to have access to recordings of individual sound sources, and assume that \textit{similar} sound sources will be present in the audio recording that we try to decompose.

As there are multiple ways to introduce prior knowledge and constraints, if given a choice, we would strongly prefer one that has the following desirable properties: it fully captures the distribution over the spectral representation of sound sources, meaning it has high modelling capacity; it is capable of modelling the underlying data generating process to some extent, and hence flexible enough to extrapolate to unseen data; it fits well into the existing NMF framework, meaning that sound sources can conveniently be added or removed from the dictionary $\mathbf{W}$.

The approach we are proposing has all of these desirable properties. Before we go on and discuss them in detail, we briefly review two much more basic ways of inferring the necessary prior information from an appropriate, additional dataset. One simple way to learn about the dictionary elements a priori, is to compute the mean spectral representation of individual sound sources. This is done by averaging over the individual frames of the spectrogram obtained from sources. Sometimes, one can already obtain reasonable decompositions with this straightforward technique. Another approach is to incorporate \textit{all} the individual spectrogram frames for a sound source directly into an \textit{overcomplete} dictionary. One can then take the sum of activations of these bases as an indicator for the presence of the sound source at a particular point in time, at the cost of additional computation.

Both of these methods have shortcomings. Simple averaging over example frames is too simplistic in most cases, and cannot adequately model realistic, high dimensional distributions. Working with overcomplete dictionaries becomes cumbersome quickly, due to both runtime and memory complexities of the decomposition, which directly depend on the number of dictionary entries --- the dictionary can not grow indefinitely in general. Both methods are still \textit{linear} and have difficulties generalizing to \textit{unseen} data.

\section{Proposed Method}
We propose a novel, flexible and principled way to incorporate prior information about the spectral characteristics of individual sound sources into the non-negative matrix factorization framework. As in supervised NMF, we assume we have access to recordings of individual sound sources that are sufficiently similar to the ones that will appear in the actual signals we want to decompose. For each of these sources, we train a \textit{normalizing flow} \cite{dinh_2017} that is capable of modelling the density of the spectrogram frames of this source. Instead of a fixed vector or a set of fixed vectors, to describe a sound source, we now have a parametrized density estimator, a kind of \textit{differentiable dictionary} at our disposal. At decomposition time, we use a collection of these differentiable dictionaries to search for a mixture of spectral representations of the sound sources that simultaneously \textit{minimizes reconstruction error} on the mixture, while \textit{staying likely} with respect to the density of the spectrogram frames of the individual sound sources.

This approach enables us to decompose an input audio mixture into linear combinations of sound sources that are best described by nonlinear processes. One scenario with these characteristics is the decomposition of piano recordings into individual notes. The mixing process in a piano is predominantly linear \cite{piano_soundboard_linear}, whereas the sound generating physical process is not \cite{piano_nonlinear}. We will use this scenario as our testbed to characterise the modelling capacity of the method.

\subsection{Nonlinear Extrapolation}
\begin{figure}[t!]
	\centering
	\includegraphics[width=\columnwidth]{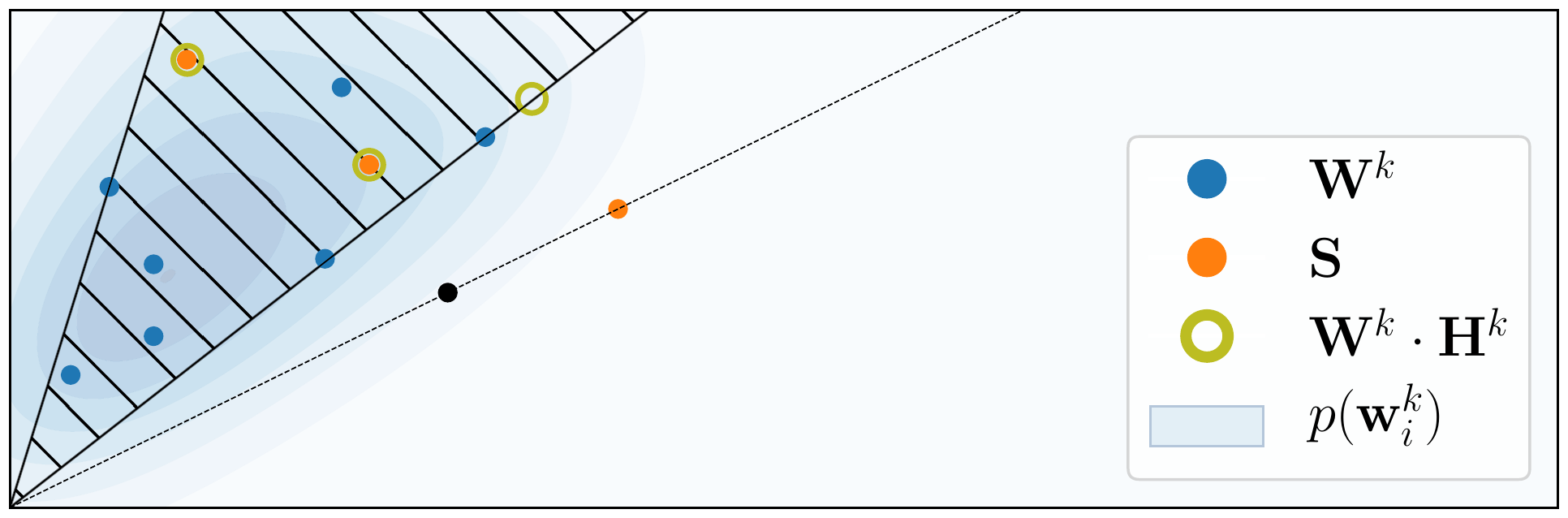}
	\caption{A simplified two dimensional sketch to illustrate nonlinear extrapolation enabled by normalizing flows.}
	\label{fig:cone_example}
\end{figure}

We will illustrate the difference between \textit{linear interpolation} with an overcomplete dictionary and \textit{nonlinear extrapolation} utilizing normalizing flows with the help of the sketch shown in Figure \ref{fig:cone_example}. Please note that this is a severely simplified example in two dimensions only, to visually support the description. The blue dots represent single feature vectors $\mathbf{w}_i^k$ from the training set that describe a particular sound source with index $k$, with $\mathbf{W}^{k}$ denoting the sub-matrix of the full dictionary $\mathbf{W}$ that contains only these feature vectors.
Due to the non-negativity constraints of NMF, these vectors form a cone $\mathcal{C} = \{ \mathbf{c} | \mathbf{c} = \mathbf{W}^{k}\mathbf{h}^{k}, \mathbf{W}^{k} \succcurlyeq 0, \mathbf{h}^{k} \succcurlyeq 0 \}$.
To reconstruct the three spectrogram frames $\mathbf{S}$ represented as orange dots in terms of this cone, and to \textit{associate} them with a sound source, they need to lie \textit{inside} this cone. This is the case for the orange dots inside the olive circles. Datapoints that lie \textit{outside} this cone cannot be reconstructed as a non-negative linear combination of the sound source feature vectors without reconstruction error. As an unfortunate secondary effect, the strength of association with this source is diminished as well.

In contrast, a normalizing flow estimates the density of the feature vectors of the sound source. It models how the feature vectors were generated, it can easily \textit{generate new samples} from this density, and is capable of \textit{nonlinearly extrapolating} to unseen new feature vectors. An example of such an unseen feature vector is shown as a black dot. It is \textit{still likely} under the density $p(\mathbf{w}_i^k)$ that is modeled by the flow, which is indicated by the blue contours. The extrapolation capabilities of the flow enable us to fully associate new, unseen, yet similar feature vectors to a given sound source, with high likelihood and almost zero reconstruction error.

\subsection{Related Work}
The concept of normalizing flows has been introduced in \cite{tabak_2010}. The particular kind of normalizing flow that we use is called \textit{RealNVP}, and has been introduced as a parametric density model in \cite{dinh_2017}. Although Generative Adverserial Networks have been used in the context of audio decomposition \cite{DBLP:conf/icassp/SubakanS18, DBLP:conf/mlsp/0063KM20, DBLP:conf/icassp/StollerED18, DBLP:conf/acssc/TanabeIFI19} they do not allow explicit access to the likelihood of a data sample. The likelihood of a data sample under a Variational Autoencoder is cumbersome to approximate as well, and necessitates Monte Carlo approximation of an expectation of a lower bound on the likelihood \cite{Bando_2018}.

\section{\method}
We introduce a novel, constrained dictionary adaptation method that we call \textit{\method{}} (\abbrev{}). This method was devised to address the problems outlined in the introduction, when decomposing mostly linear mixtures of non-linearly behaving sound sources.

\subsection{The \abbrev{} Model}
We denote the magnitude spectrogram that we would like to decompose as $\mathbf{S} \in \mathbb{R}_{+}^{D \times T}$, having $D$ spectral bins and $T$ time frames. We denote the fixed dictionary with $N$ entries as $\mathbf{W} \in \mathbb{R}^{D \times N}$, and the activation matrix as $\mathbf{H} \in \mathbb{R}^{N \times T}$. Supervised NMF seeks to approximate the spectral frame $\mathbf{s}_{t} \approx \sum_{n} h_{t}^{n} \mathbf{w}^{n}$, where $h_{t}^{n}$ is the activation of the $n$-th dictionary entry $\mathbf{w}^{n}$ at time $t$. We will now describe how to integrate \abbrev{} with the NMF framework.

To estimate the density of the spectrogram frames for each sound source, we use a simplified version of RealNVP \cite{dinh_2017} without multi-scale architecture, batch normalization, or spatial masking. Instead, we use plain multi-layer perceptrons to parametrize the affine coupling layers, and randomly chosen, fixed permutation layers in between the coupling layers.

For each of $K$ sound sources, we train a separate normalizing flow to obtain a parametrized density estimator $f_{k}$. A flow allows explicit evaluation of the likelihood $p(\mathbf{x})$ of a sample $\mathbf{x}$ by computing $p_Z(f_{k}(\mathbf{x}))$. A flow can generate samples by drawing $\mathbf{z} \sim p_Z$ and computing $f_{k}^{-1}(\mathbf{z})$, where $p_Z(\mathbf{z})$ is a prior distribution, which we choose to be a standard isotropic Gaussian $\mathcal{N}(\mathbf{\mu} = \mathbf{0}, \, \mathbf{\Sigma} = \mathbf{\mathit{I}})$, following \cite{dinh_2017}.

\abbrev{} approximates an arbitrary spectrogram frame $\mathbf{s}_{t}$ as a weighted sum of dictionary components $\mathbf{s}_{t} \approx \hat{\mathbf{s}}_{t} = \sum_{k} h_{t}^{k} \mathbf{w}_{t}^{k}$ of $K$ sound sources. The $k$-th component is generated by the $k$-th flow as $\mathbf{w}_{t}^{k} = f_{k}^{-1}(\mathbf{z}_{t}^{k})$. For decomposition, \abbrev{} updates the component activations $\mathbf{h}_{t} \in \mathbb{R}_{+}^{K}$ and the dictionary entries \textit{in latent space} $\mathbf{Z}_{t} \in \mathbb{R}^{D \times K}$ that \textit{generate components in data space} via the flows $\{f_{k}\}_{k=1}^{K}$, using (projected) gradient descent on the loss $\mathcal{L}$. Minimizing the loss jointly minimizes reconstruction error of the mixture, and maximizes likelihoods of individual components:

\begin{equation}
\mathcal{L}(\mathbf{s}_{t}, \hat{\mathbf{s}}_{t}) = \| \mathbf{s}_{t} - \hat{\mathbf{s}}_{t} \|_2 - \frac{c}{D \sum_{k} h_{t}^{k}} \sum_{k} h_{t}^{k} \log p_Z(\mathbf{z}_{t}^{k})
\label{eq:dds_loss}
\end{equation}

\begin{figure}[t!]
	\centering
	\includegraphics[width=\columnwidth]{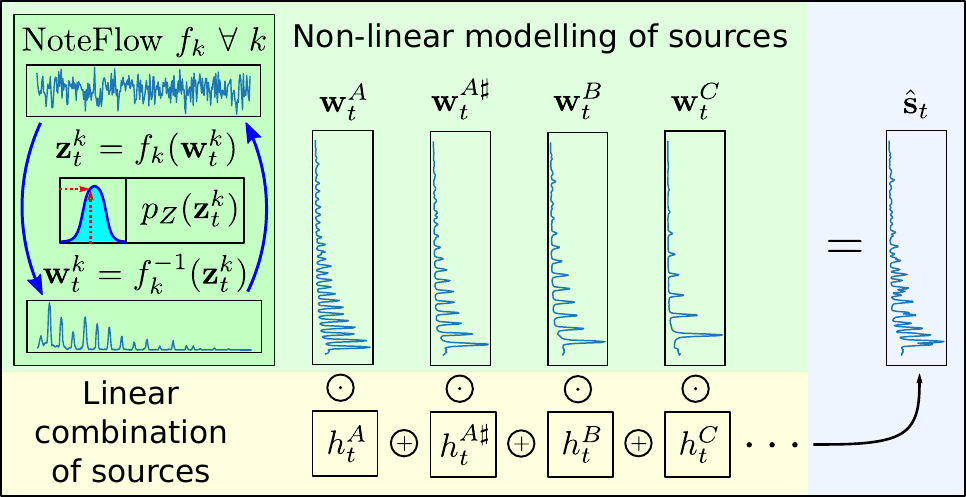}
	\caption{An illustration of \abbrev{} decomposing a spectrogram frame.}
	\label{fig:method_principle}
\end{figure}

The likelihood penalties $- \log p_Z(\mathbf{z}_{t}^{k})$ on the latent vectors of individual sources are normalized to \textit{nats per dimension} by $\frac{1}{D}$ and weighted by the activation components $h_{t}^{k}$ normalized by their sum $\sum_{k} h_{t}^{k}$. The global likelihood penalty weight $c$ is a hyperparameter of the decomposition and allows to fine-tune the behavior of the method, balancing reconstruction quality and deviations from \textit{likely} dictionary entries. We provide an illustration of the process in Figure \ref{fig:method_principle}.

\subsection{Expected Benefits and Open Questions}
Our design addresses the performance limitations of the strictly linear, overcomplete baseline when we are dealing with non-linearity of sound sources, while preserving interpretability. Due to \abbrev{} modelling the data generating process, the ability to represent arbitrary mixtures of similar sound sources should be considerably improved. Another benefit is the ability to better cope with the inevitable \textit{distribution shift} between the data used for training the individual sound source models, and the data which is encountered during decomposition, while preserving association strength with a sound source.

To investigate the usefulness of the added non-linear extrapolation capability, we designed a set of highly controlled experiments aimed at revealing the differences between overcomplete NMF and \abbrev{}.
Both methods need to approximate both the mixing and the reconstruction aspects of the signal-generating process. Both assume a linear mixing process; the approximate reconstruction of sound sources is entirely different, however.
Thus, we will first study and compare the \textit{reconstruction capabilities}. We quantitatively evaluate the differences in ability to generalize to new, unseen data for isolated sound sources in Section \ref{sec:nonlinear_extrapolation_eval}. This gives us a first appraisal of the beneficial aspects of \abbrev{}, before we compare the decomposition capabilities. We ensured equal train and test conditions for both methods, for a fair comparison.

Similarly, we would like to study the \textit{discrimination ability} of individual NoteFlows in isolation. NoteFlows need to assign high likelihoods to unseen samples that are similar to the training samples, and low likelihoods to unseen and dissimilar samples. This is a necessary prerequisite for a clear decomposition into pre-defined components. We quantify the one-vs-all discrimination ability of the individual NoteFlows, by determining \textit{likelihood thresholds}. These permit both soft and hard constraints on what samples can be generated by the sound source models, and hence facilitate calibration of the method. They act as a kind of \textit{natural similarity measure} to the data generating process: we can minimize misattributions by constraining generated samples to have likelihoods below these thresholds. We will quantify one-vs-all discrimination on a set of models in Section \ref{sec:crosseval_flows}. The complete \abbrev{} decomposition scheme will be put to a practical test in \ref{sec:semantic_decomposition_test}.

\section{Experimental Evaluation}
\label{sec:experimental_evaluation}
In order to compare how the methods deal with \textit{distribution shift} from sound sources seen during training to (similar) sound sources encountered during testing, we extract a set of notes played on various acoustic piano instruments, with different volume levels (also called ``velocity'', in MIDI jargon) and varying acoustics conditions. We divide the isolated notes into multiple splits consisting of two disjoint subsets: a training dataset and a test dataset. We draw upon a commercial VSTi sample-based synthesizer plugin, called ``Spectrasonics KeyScape''\footnote{\url{https://www.spectrasonics.net/products/keyscape/}}, which contains a multitude of high quality samples of isolated notes played on different acoustic pianos in various microphone conditions and room reverberation settings, that provide a rich variety of timbre.

We sample the 4 notes A1, A2, A3, A4 and an additional full octave A2-A3 (12 notes). For each isolated note, we extract audio recordings played with 4 representative velocity values (32, 64, 96, 127) on \textit{all} of the 43 available acoustic piano presets. We create 9 different splits by using different subsets of presets for training and testing, to create multiple distribution shift scenarios that challenge both methods, and serve as a realistic testbed displaying real-world characteristics.

We downsample the isolated note recordings to $16$ [kHz] mono waveforms and compute logarithmic magnitude spectrograms with a Hann window of size $2048$ and hop size $512$. We only keep the lowest $512$ resulting spectral bins that represent the frequency range from $[0; 4]$ [kHz], and normalize the magnitudes to the interval $[ 0; 1]$. Depending on the particular split, we have $N$ spectrogram frames as $512$-dimensional training examples, where $N$ lies in the range $[3504, 25228]$.

To evaluate \abbrev{}, we train a RealNVP model for each isolated note in the training set. The models have 16 affine coupling layers, and each coupling function is realized as a multi-layer perceptron with 4 dense layers of 256 units with SELU activations~\cite{DBLP:conf/nips/KlambauerUMH17}. The Adam optimizer~\cite{Kingma2015} is used with a learning rate of $1 \cdot 10^{-3}$ to train each model for up to $1000$ epochs, with a minibatch size of 512. The early stopping criterion was configured to have a patience of $50$ epochs. Density estimation performance is monitored by computing the mean log-likelihood over a held-out validation set. The validation set consists of 20\% randomly selected training samples. We refer to these models, trained on isolated notes, as ``NoteFlows'' throughout this manuscript. The parameters of each NoteFlow are fixed during decomposition.

We compare \abbrev{} to an overcomplete variant of NMF, with its dictionary $\mathbf{W}$ initialized to the training set and held fixed. The activation matrix $\mathbf{H}$ is initialized to constant values of $\frac{1}{N}$ where $N$ is number of components given by size of the training set.

NoteFlows generate spectrogram frames determined by the noise vectors $\mathbf{z}_t^k$. We initialize these noise vectors to $\mathbf{0}$ and set the global likelihood penalty weight
(see Eq.\ref{eq:dds_loss}) to $c = 1 \cdot 10^{-3}$ for all experiments that follow.

The decomposition is run for a maximum of 10000 update steps for both methods, overcomplete NMF and \abbrev{}. Early stopping is possible, as soon as the cost term changes by less than $\epsilon = 1 \cdot 10^{-15}$, between update steps. If the cost term fluctuates, but there is no progress for $10$ consecutive steps, the learning rate is reduced by a factor of $2$. For all decompositions, the Adam~\cite{Kingma2015} optimizer is used. After each update step, the current solution is projected back into the non-negative orthant to satisfy all constraints.

\subsection{Non-linear Extrapolation}
\label{sec:nonlinear_extrapolation_eval}

\begin{figure}[t!]
	\centering
	\includegraphics[width=\columnwidth]{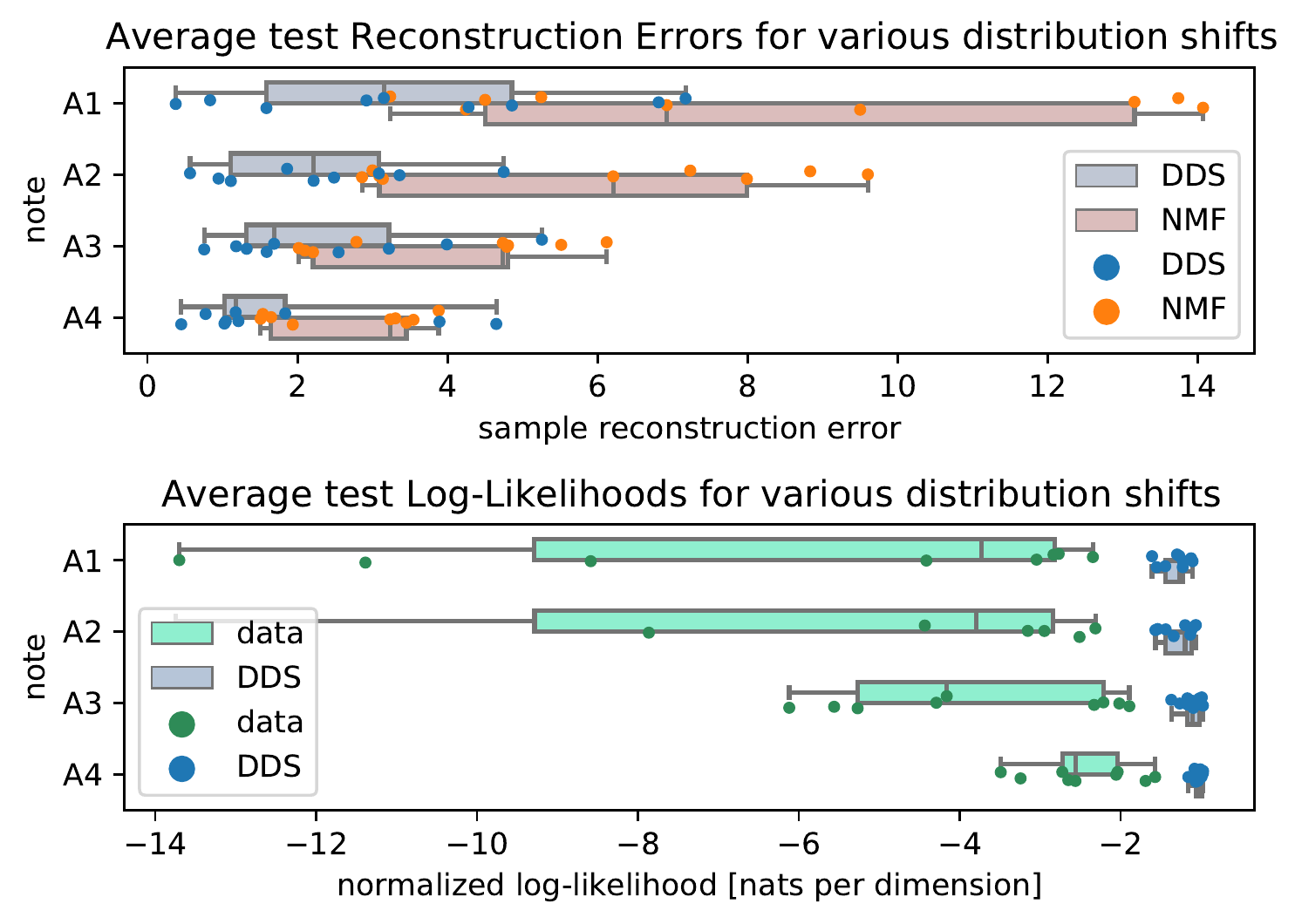}
	\caption{Quantitative assessment of robustness to distribution shift of the compared methods \abbrev{} and overcomplete NMF.}
	\label{fig:dds_vs_nmf_quantitative}
\end{figure}

After \abbrev{} is trained using the aforementioned procedure, and the dictionary matrix of overcomplete NMF is initialized with the spectrogram frames of the same train set, we study the reconstruction error of each method on new, unseen sound sources. This is shown in the upper half of Figure \ref{fig:dds_vs_nmf_quantitative}, where each data point represents the average reconstruction error over the test samples of a particular split. In the lower half of Figure \ref{fig:dds_vs_nmf_quantitative} we look at the tradeoff between reconstruction error and sample likelihood,. The likelihoods of test samples with zero reconstruction error (labeled as \texttt{data}) are contrasted with the likelihoods of test samples where a small amount of reconstruction error is allowed, while the sample is kept likely (labeled as \texttt{DDS}).

We can see clear evidence for the advantage of \abbrev{} over the overcomplete, linear NMF baseline when it comes to reconstruction quality. This is especially apparent for the low note A1, but can be observed across all four octaves. The lower half of Figure \ref{fig:dds_vs_nmf_quantitative} shows a noticable decrease in likelihood if one insists on a perfect reconstruction. What we can observe is that \abbrev{} is perfectly capable of producing reconstructions with \textit{low error} while still staying \textit{close} to the training data distribution.

\subsection{Discrimination Ability of NoteFlows}
\label{sec:crosseval_flows}

\begin{figure}[t!]
	\centering
	\includegraphics[width=.8\columnwidth]{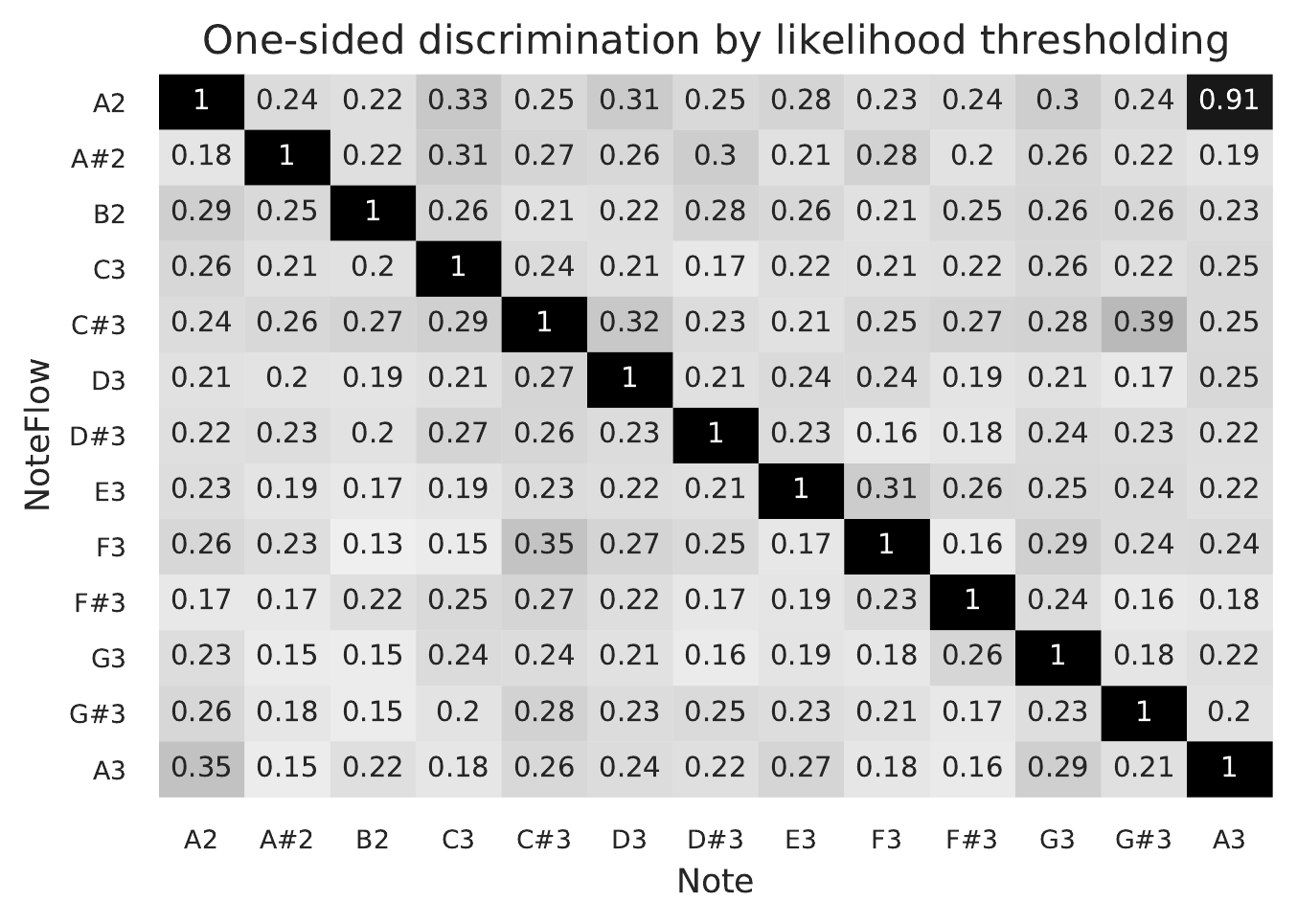}
	\caption{Confusion of NoteFlows in terms of likelihood-based one-sided discrimination ability between their ``correct'' notes, and all the other notes.}
	\label{fig:osd_conf_mat}
\end{figure}

\begin{figure}[t!]
	\centering
	\includegraphics[width=\columnwidth]{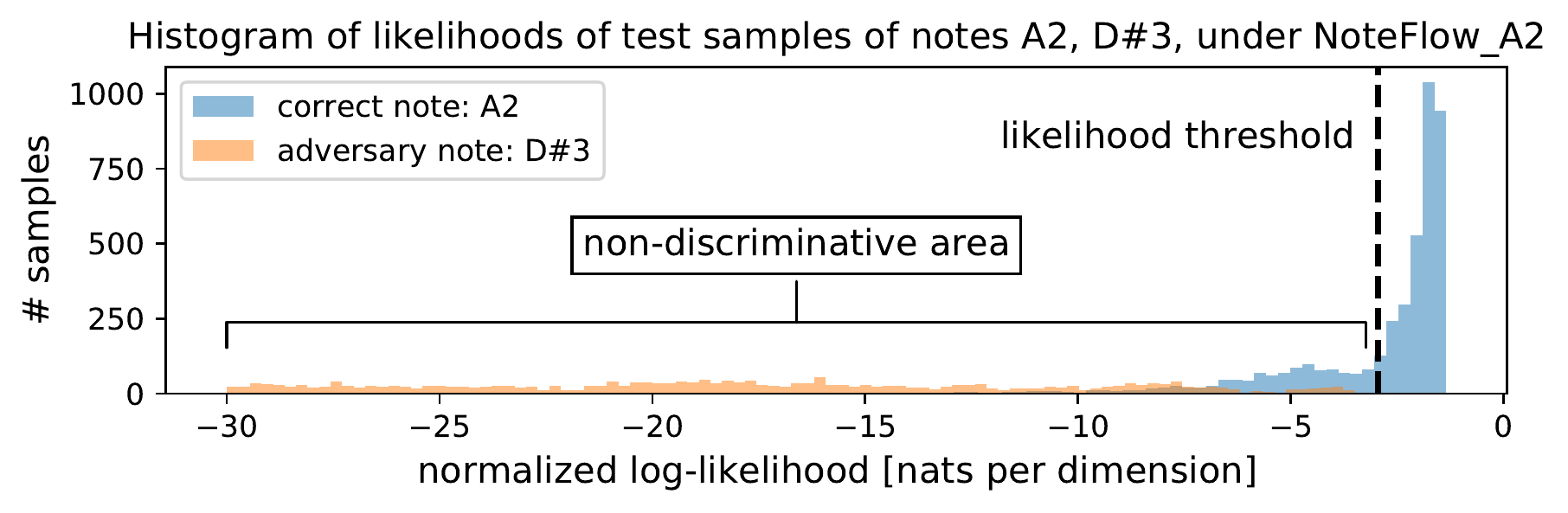}
	\caption{Illustration of likelihood thresholding in definition of one-sided discriminativeness quantifier.}
	\label{fig:likelihood_overlap}
\end{figure}
\begin{figure*}[t!]
	\centering
	\includegraphics[width=\textwidth]{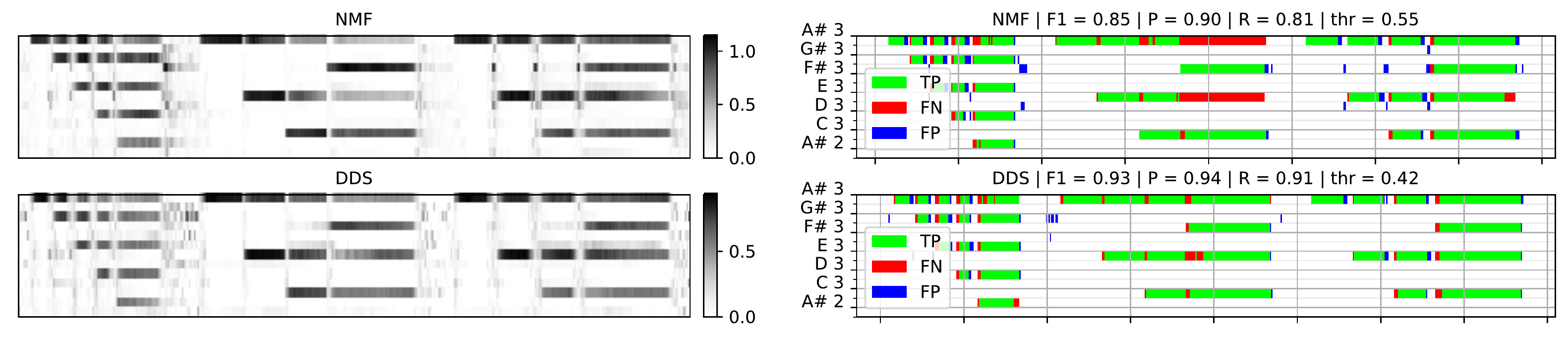}
	\caption{Demonstration of semantic decomposition capability. Activation matrices; on the left. Thresholded activation matrices visualizing the underlying computation of the (M)IR metrics by highlighting where True Positives (TP), False Negatives (FN) and False Positives (FP) are; on the right.}
	\label{fig:semantic_decomposition}
\end{figure*}

To assess the potential for \abbrev{} to wrongly attribute spectral activity, we measure \textit{one-sided discriminativeness} on one representative split. Given a model of source $k$ and a set of $N$ test samples $\mathbf{x}^k$ similar to source $k$ as well as test samples $\mathbf{x}^j$ of a different source, the one-sided discriminativeness measure is defined as the ratio of the number of samples of source $k$ (blue histogram in Figure \ref{fig:likelihood_overlap}) below the ``likelihood threshold'', to the number of \textit{all} samples of source $k$ (entirety of blue area in Figure \ref{fig:likelihood_overlap}). The likelihood threshold $\theta$ is defined as the \textit{maximum} over the likelihoods of the samples of the \textit{different} source $j$ under the NoteFlow model for source $k$. This leads to the expression $d_{os} = \left( \sum_{n=1}^{N} [p(\mathbf{x}_{n}^{k}) < \theta] \right) / N$ and, intuitively speaking, measures how many of the samples that the NoteFlow should model as ``likely'' -- using likelihood as a kind of natural similarity measure -- are confused with samples of some other source, that should be deemed ``unlikely'' under this particular NoteFlow.

We trained 13 NoteFlows to model one full octave of piano notes, and computed the one-sided discriminative measure for all of them, using unseen data from the test set. The entries in the confusion matrix in Figure \ref{fig:osd_conf_mat} are the ratios as previously defined. Lower numbers are better. The main diagonal is always one by definition. Out of curiosity, we computed the one-sided discriminativeness with samples solely from the training data, which yielded a confusion matrix with \textit{all} entries very close to zero. This shows that the method is capable of explaining the majority of spectral evidence by assigning correct sources. We find this encouraging, and note that there is still room for improvement regarding the generalization capabilities of the parametric density estimation models we currently employ.

\subsection{Semantic Decomposition}
\label{sec:semantic_decomposition_test}

Finally, we compare the abilities of \abbrev{} and overcomplete NMF to decompose a short piece of polyphonic piano music into its individual notes. We rendered the piece twice, once with an instrument seen during training, and once with a new, unseen instrument, using the same splits as in Section \ref{sec:crosseval_flows}. Each method needs a global threshold to binarize the activity matrix $\textbf{H}$, so the F1-score for frame-level source attribution can be computed. The optimal threshold for each method is a trainable parameter, and is determined on the training piece. During test time the two respective thresholds are held fixed and used to binarize the activity matrices of the decompositions. Finally, we compute the F1-score for the binary activity matrices on the test piece.

The results of these decompositions are shown in Figure \ref{fig:semantic_decomposition}. The left column shows the raw activations of source components for each method. For the overcomplete NMF approach, each row of the activation matrix is computed as the sum of the activities of all dictionary entries belonging to a given source. The right column shows the binarized activations, contrasted with the ground truth. The resulting F1-score can be found in the titles.

\section{Conclusion}
We observe that despite the possibility for confusion of sources, as measured in Figure \ref{fig:osd_conf_mat}, \abbrev{} demonstrates its ability to strongly associate spectral activity to the appropriate source. This demonstrates a clear improvement of decomposition performance over the linear, overcomplete NMF baseline, as measured by the frame-level F1 metric. The improvement can be directly attributed to the great improvement in recall, and we interpret this improvement as direct evidence for the usefulness of introducing non-linear extrapolation capability into the general NMF framework to jointly achieve clearer decompositions and better reconstructions of the sources.

\bibliographystyle{IEEEtran}
\bibliography{main}

\end{document}